\documentclass[a4paper,12pt]{revtex4}
\usepackage{amssymb}
\usepackage{amsmath}
\usepackage{graphics}
\usepackage{color}
\usepackage[T1]{fontenc}
\usepackage{times,mathptm}
\usepackage[cp1250]{inputenc}
\usepackage{epsfig,epsf}
\newcommand{\be}{\begin{equation}}
\newcommand{\ee}{\end{equation}}

\newcommand\beq{\begin{eqnarray}}
\newcommand\eeq{\end{eqnarray}}
\newcommand\nvec[1]{\textbf{\emph{#1}}}

\begin{document}

\title{On relevance of triple gluon fusion in $J/\psi$ hadroproduction}

\author{Leszek Motyka}
\email{leszek.motyka@uj.edu.pl}
\author{Mariusz Sadzikowski}
\email{mariusz.sadzikowski@uj.edu.pl}
% \email{leszek.motyka@desy.de}
\affiliation{Institute of Physics, Jagiellonian University, \L{}ojasiewicza 11, 30-348, Krak\'{o}w, Poland
}

\date{January 21, 2015}

\begin{abstract}
A contribution to $J/\psi$ hadroproduction is analyzed in which the meson production is mediated by three-gluon partonic state, with two gluons coming from the target and one gluon from the projectile. This mechanism involves double gluon density in one of the protons, hence this contribution enters at a non-leading twist. It is, however, relevant due to an enhancement factor coming from large double gluon density at small~$x$. We calculate the three-gluon contribution to $J/\psi$ hadroproduction within perturbative QCD in the $k_T$-factorization framework. Results are obtained for differential $p_T$-dependent cross-sections for all $J/\psi$ polarizations and for the sum over the polarization components. The rescattering contribution is found to provide a significant correction to the standard leading twist cross-section at the energies of the Tevatron or the LHC at moderate $p_T$. We suggest $J/\psi$ production in proton-nucleus collision as a possible probe of the triple gluon mechanism.
\end{abstract}

\maketitle

\section{Introduction}

Production of heavy vector quarkonia in proton collisions has been a subject of intense experimental and theoretical investigations since CDF found huge excess of the measured cross-sections of prompt $J/\psi$~\cite{TeV0} with respect to the standard theoretical predictions in QCD. Initially the theory predictions were based on a leading order collinear QCD approximation, in which the hard matrix element of a partonic color singlet subprocess $gg \to J/\psi g$ gives the dominant contribution. This approximation, however, underestimates badly the measured cross-section and yields an incorrect $p_T$ distribution of the produced mesons, so it was necessary to consider alternative mechanisms of quarkonia hadroproduction. In particular, two theoretical concepts received a lot of attention: the so called color octet model (COM) \cite{Bodwin,CL1,CL2,BK1,BK-NLOfit,BK-pol,MaWaCha,Chao} and the other approach based on the $k_T$-factorization (KTF) scheme \cite{GLR,HKSST,Baranov1,Baranov2,Baranov3,KVS}. For a comprehensive theory review of various approaches to quarkonium hadroproduction see \cite{Lansberg}.

The COM is based on the fact that the hadron wave functions contain higher Fock components. For heavy quarkonia it implies that the heavy quark-antiquark ($Q\bar Q$) pair in a heavy meson may be accompanied e.g.\ by one or more gluons. So the $Q\bar Q$-pair may appear in the meson in a color octet state \cite{Bodwin}. Amplitudes of such octet components are subleading (w.r.t.\ the standard, color singlet component) in heavy quark velocity expansion within the Heavy Quark Effective Theory \cite{Bodwin}, but in the vector meson hadroproduction the color octet states contributions are enhanced by the corresponding hard partonic cross-section for production of such states \cite{CL1,CL2}. With only a few free parameters the color octet model is quite successful in describing the magnitude and $p_T$-dependence of the prompt $J/\psi$ hadroproduction \cite{BK-NLOfit,BK-pol,MaWaCha,Chao}, however, it faces some problems with an accurate desription of the produced mesons polarization. In particular, predictions for the meson polarization change a lot when going from LO to NLO accuracy \cite{BK-pol,Chao}. Besides, neither LO nor NLO COM estimates of prompt $J/\psi$ polarization are fully consistent with the data. It is fair to add that in the COM the amplitudes of color octet $Q\bar Q$ transitions into mesons are not known from the first principles, and the theoretical estimates of the cross-section magnitudes rely upon fitted parameters. On the other hand, kinematical dependence of the cross-sections, e.g.\ the $p_T$-dependence, is already a genuine QCD prediction (for each partonic subprocess separately) and the good description of the meson $p_T$-dependence within the COM supports strongly validity of this approach. The COM may be also understood within hard factorization theorem in QCD, where universal (independent on the environment) fragmentation functions exist of $Q\bar Q$ partonic states with various quantum numbers (including color octet) into final state quarkonia \cite{KQS}.

Another theoretical approach to quarkonia hadroproduction relies on the $k_T$-factorization scheme. In this framework, one allows for non-zero transverse momentum of colliding partons. Because of multiple emissions in the course of QCD evolution the parton momentum may become sizeable if the evolution length is sufficiently large. This effect is especially pronounced if gluons are probed at small~$x$, which is the case for prompt quarkonia production at the Tevatron or at the LHC. Then the kinematical distributions of produced quarkonia may be strongly affected by $k_T$ of incoming gluons. Indeed, inclusion of gluon $k_T$ in the analysis brings theoretical estimates much closer to the experimental data, both for the magnitude and for the shape of the distributions, already at the lowest order in QCD. This works rather well for the $C$-even charmonia, like $\chi_c$, but for the accurate description of Tevatron data on prompt $J/\psi$ and $\psi'$ production a color octet component was necessary also in the KTF approach \cite{Baranov1}, although, smaller than in the collinear approximation. For the recent LHC data on prompt charmonia production, however, a good theoretical description of magnitude, kinematics and polarization was found within KTF without including the color octet mechanism \cite{Baranov3}. This good description in KTF CSM was achieved by using a more advanced unintegrated parton density, CCFMA0 \cite{ccfma0}. Still, it is fair to say that a global, consistent description of quarkonia production in KTF has not been formulated yet.

It follows from this discussion that theoretical understanding of heavy quarkonia hadroproduction is not yet fully satisfactory. Polarization description and the need of fitting key parameters are somewhat weak points of the color octet model, and a global description of $J/\psi$, $\psi'$ and $\chi_c$ hardroproduction was not formulated yet in the KTF approach. Therefore it is possible, that both proposed approaches are not complete and need to be supplemented by another mechanism.

The COM and CSM approaches both in the collinear and KTF framework share an important common feature --- it is assumed that at the parton level the quarkonia production is initiated by two gluons. At very large energies, however, when the gluon density becomes large, it is expected that sizable corrections to this picture may come from subprocess with more than two initial state partons, which would be classified as rescattering or shadowing/anti-shadowing contributions. Such effects are expected to be particularly important in collisions with heavier nuclei. First estimates of such multiple-gluon contributions to heavy quarkonia production in $pp$ and $p\bar p$ collisions were performed by Khoze, Martin, Ryskin and Stirling \cite{KMRS}, who considered three-gluon initial partonic state. Recently, Ma and Venugopolan \cite{MV} proposed resummation of rescattering contributions within the color glass condensate (CGC)  formalism \cite{cgc1,cgc2,cgc3}. A good description of unpolarized data for $J/\psi$ and $\psi'$ was achieved down to very low $p_T$ by matching the resummed CGC cross-section with predictions of the collinear NRQCD COM approach \cite{MV,KMV}.

In this paper, following the idea of Khoze, Martin, Ryskin and Stirling \cite{KMRS}, we shall investigate the three gluon mechanism of quarkonia hadroproduction in which hard rescattering plays the key r\^{o}le in color neutralization of the $Q\bar Q$ pair. Thus, we shall consider the 3-gluon fusion process at the parton level: $3g \to J/ \psi$ in the color singlet channel instead of the standard color singlet process $gg \to J/\psi g$. This three gluon fusion partonic process is, of course, a higher twist correction to the cross-section, as it couples to a double gluon density in one of the protons. Therefore, one expects it to be suppressed w.r.t.\ the standard, leading twist contribution by powers of the hard scale. In high energy collisions, however, this rescattering term receives significant enhancement due to large gluon density, reflected by large ratio of double and single gluon density at small-$x$. Moreover, the matrix elements of the rescattering term $3g \to J/ \psi$, and the standard one $gg \to J/\psi g$, come at the same order of perturbation theory, so no additional $\alpha_s$ suppression of the rescattering piece occurs in the matrix element. Hence, despite of the higher twist nature, it may be still important. As already stated, such processes were already proposed in Ref.~\cite{KMRS} where estimates were given showing that rescattering might explain the discrepancy between the data and the predictions based on the color singlet mechanism. We address and develop this idea in more detail, by performing a complete calculation of the rescattering contribution in the KTF framework, including an explicit calculation of $p_T$ distribution of produced quarkonia and the polarization composition. We find that the rescattering correction is a significant contribution to the prompt $J/\psi$ cross-section especially at moderate $p_T$ and should be included in a precise description of the data. It should be stressed that the rescattering correction complements in natural way both the COM and the standard KTF approach, providing a kinematics- and process-dependent mechanism of color neutralization. Also, we expect that the rescattering mechanism should be strongly enhanced and thus -- even more important for heavy quarkonia production in collisions with nuclei.

The paper is organized as follows: in Sec.\ \ref{sec:2} we introduce the kinematics of the process and the notation, in Sec.\ \ref{sec:3} we derive analytic formulae for the amplitudes and cross-sections of the direct quarkonium production by rescattering, in Sec.\ \ref{sec:4} we give numerical estimates for the cross sections and conclusions are given in Sec.\ \ref{sec:5}. Technical pieces of derivations are given in the Appendix.

\section{Kinematics}
\label{sec:2}

\begin{figure}
\centerline{
a) \includegraphics[width=0.32\columnwidth]{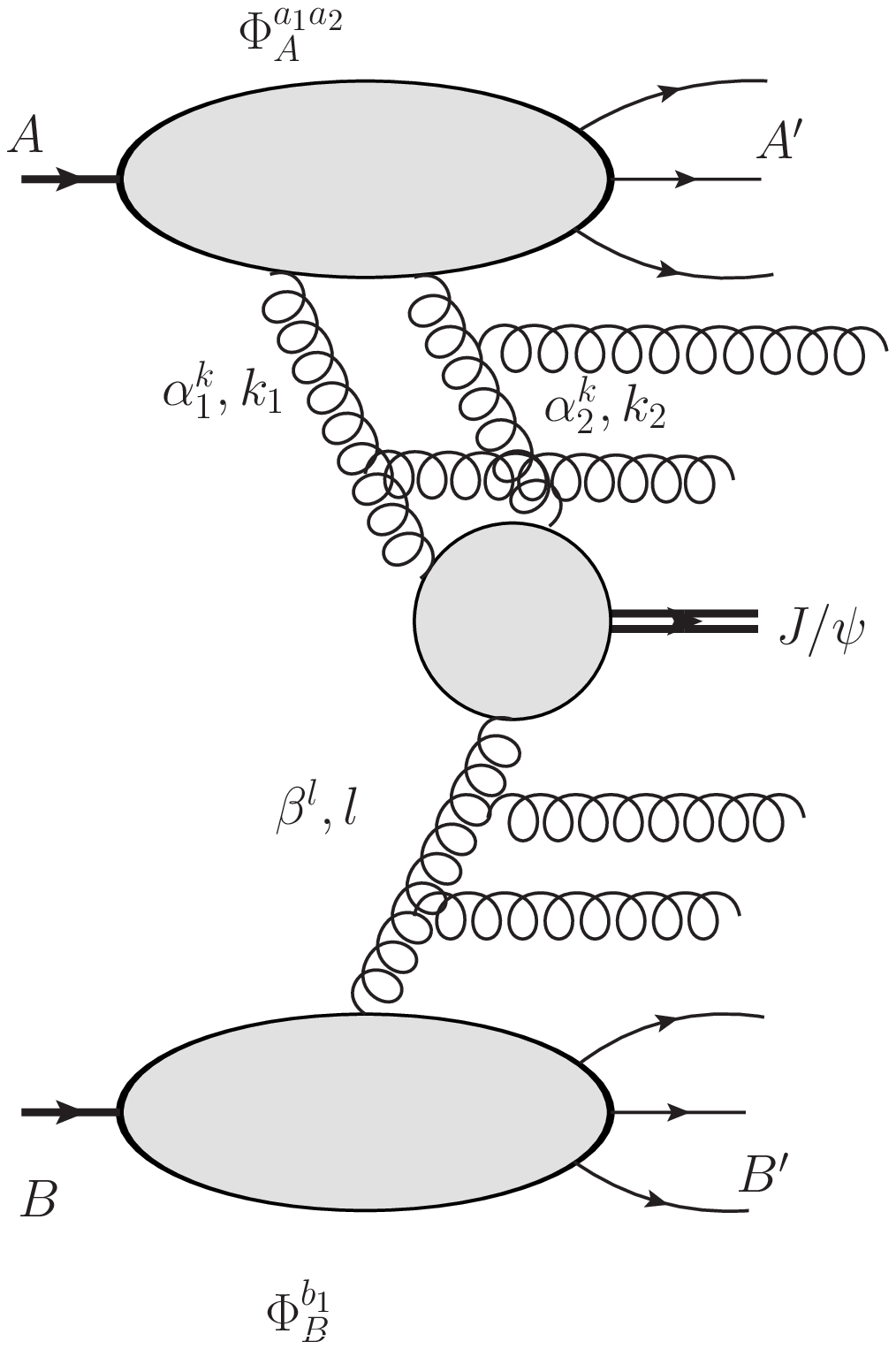}\hspace{3em}%\vspace{-1em}
b) \includegraphics[width=0.48\columnwidth]{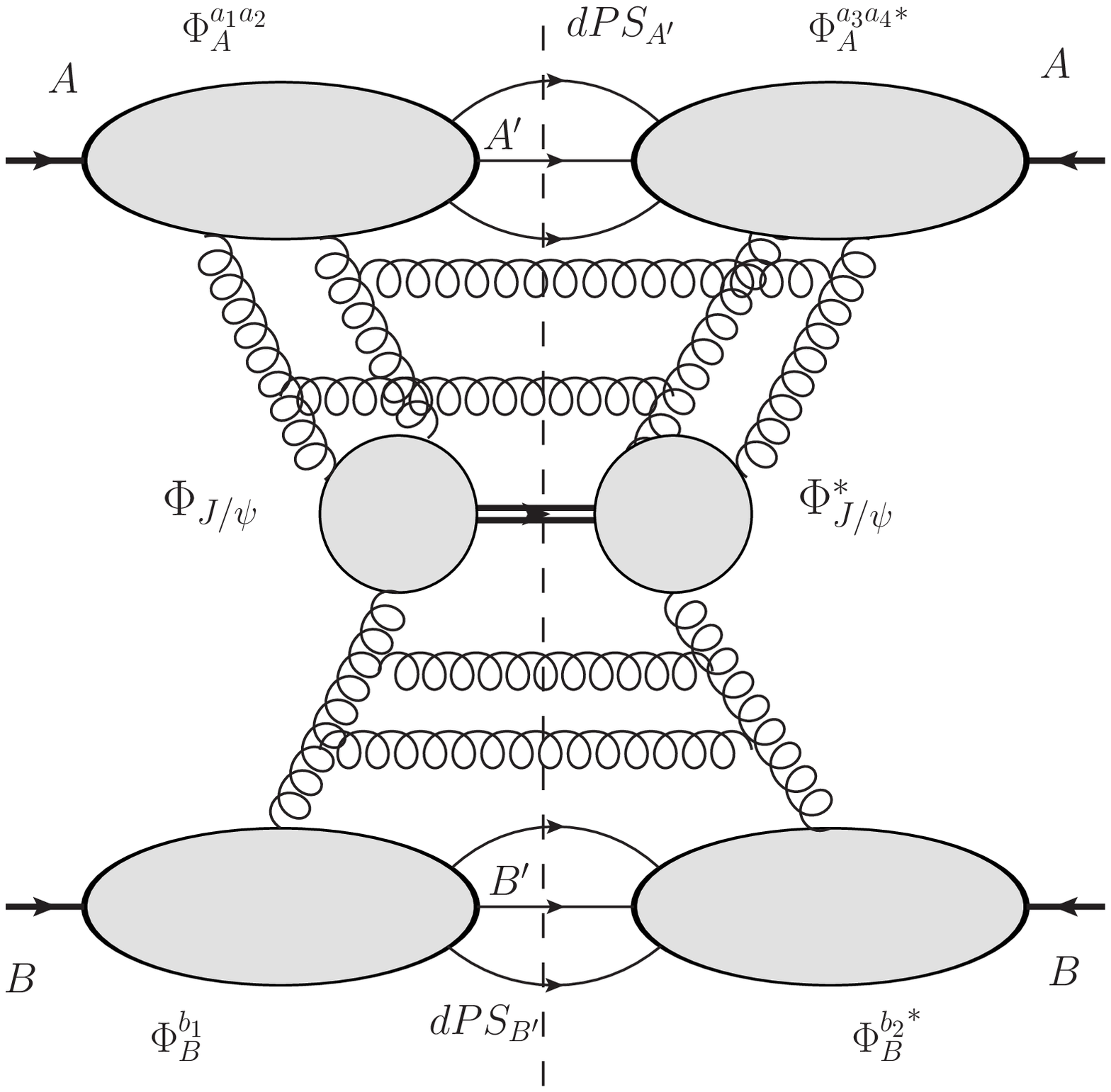}%\vspace{-1em}
}
\caption{a) The amplitude $\mathcal{M}$ of the $J/\psi$ hadroproduction in the triple gluon fusion mechanism; b) the dominant diagram in the square of the amplitude, $|\mathcal{M}|^2$.
\label{fig1}
}
\end{figure}

We consider direct inclusive production of heavy quarkonia in high energy proton--(anti-) proton collisions and focus on a process mediated by partonic subprocess with tree incoming gluons:  $3g \to V$. In general, $V$ can be any heavy vector quarkonium, $J/\psi$, $\psi'$, $\Upsilon$ and so on, but we choose the best measured $J/\psi$ production as the reference process. The three-gluon fusion contribution enters as a higher twist correction to the direct vector quarkonium production. The notation for the four-momenta of incoming protons ($p_A$, $p_B$), the outgoing meson, $p$, outgoing hadronic final states ($p'_A$ for $A'$ and $p'_B$ for $B'$) is explained in the left panel of Fig.\ \ref{fig1}.

We analyze the process within perturbative QCD in the high energy approximation, using the framework of $k_T$-factorization. In the high energy limit, only the leading power of $s= (p_A + p_B)^2$ is retained. In this approximation the proton masses may be neglected, $p_A^2=p_B^2=0$, and $s \simeq 2 p_A \cdot p_B$. The masses of outgoing states are denoted by $p^2 = M^2_{J/\psi}$, $p_{A^\prime}^2=M_{A^\prime}^2$, $p_{B^\prime}^2=M_{B^\prime}^2$ for $J/\psi$ and hadronic final states $A'$ and $B'$ emerging from the target and the projectile correspondingly.

The diagram in Fig.\ \ref{fig1}a illustrates the partonic topology of the amplitude for direct vector quarkonium production with three intermediate gluons.
%The evaluation of this diagram amplitude allows for an estimate of the rescattering correction to the direct vector quarkonium production.
The four-momenta of intermediate gluon couplings to $p_A \to A'$ and $p_B \to B'$ transitions are denoted by $k_1$, $k_2$ and $l$ correspondingly. The four-momentum conservation imposes constraints on the four-momenta,
\be
\label{kinematics1}
p_A = p_{A^\prime}+k_1+k_2,\;\;\; p_B=p_{B^\prime}+l,\;\;\; p = k_1+k_2+l.
\ee
We employ the Sudakov parameterization of the four-momenta,
\be
p_{A^\prime(B^\prime)} = \alpha_{A(B)}p_A + \beta_{A(B)}p_B + p_{A^\prime(B^\prime)\perp},\;\;\;p=\alpha p_A+\beta p_B+p_\perp \, ,
\ee
where the transverse directions are in the plane orthogonal to the direction of the collision axis in the CMS frame. Then, the kinematic constraints lead to:
\be
\beta_A = \frac{\nvec{p}^{\,2}+M_{A^\prime}^2}{\alpha_A s},\;\;\; \alpha_B = \frac{\nvec{p}^{\,2}+M_{A^\prime}^2}{\beta_B s},\;\;\;
\beta = \frac{\nvec{p}^{\,2}+M_{J/\psi}^2}{\alpha s} .
\ee
The process of $J/\psi$ production is analyzed in the kinematical domain where $\alpha_A=O(1), \beta_B=O(1)$ and
consequently $\beta_A\ \sim \alpha_B \sim O(1/s)$. Thus, the gluon momenta satisfy
\be
k \equiv k_1+k_2 \approx (1-\alpha_A)p_A-p_{A^\prime\perp},\;\;\; l \approx (1-\beta_B)p_B-p_{B^\prime\perp}
\ee
and from the last term of (\ref{kinematics1}) one gets
\be
\alpha\approx 1-\alpha_A,\;\;\; \beta \approx 1-\beta_B,\;\;\; p_{\perp} = -p_{A^\prime\perp}-p_{B^\prime\perp} .
\ee
In the Sudakov parameterization of the gluon momenta $k_i=\alpha^k_i p_A+\beta^k_i p_B+k_{i\perp}$, $l=\alpha^l p_A+\beta^l p_B+l_{\perp}$ (see Fig.\ \ref{fig1}a)
one has relations:
%\beq
%&&\alpha^k_1+\alpha^k_2 = \alpha^k_3+\alpha^k_4 = \alpha, \;\; \beta^k_1+\beta^k_2=\beta^k_3+\beta^k_4 = O(1/s),\;\;
%\alpha^l_1=\alpha^l_2 = O(1/s), \;\; \beta^l_1=\beta^l_2=\beta,\;\; \nonumber\\
%&&\nvec{k}_{1 }+\nvec{k}_{2 } = \nvec{k}_{3 } + \nvec{k}_{4 } \equiv \nvec{k},\;\;
%\nvec{k}_{ }+\nvec{l}_{1 } = \nvec{k}_{ }+\nvec{l}_{2 } = \nvec{p}_  .
%\eeq
\beq
\alpha^k_1+\alpha^k_2 = \alpha, \;\; \beta^k_1+\beta^k_2 = O(1/s),\;\;
\alpha^l = O(1/s), \;\; \beta_1=\beta.
\eeq

\section{The triple gluon contribution: analytic formulae}
\label{sec:3}

\subsection{The triple gluon amplitude}

The amplitude in Fig.\ \ref{fig1}a reads
\be
\label{amplitude1}
-i\mathcal{M} = \frac{1}{2!} S^{\,b_1}_{B\,\mu_1}(p_B,r_B,B^\prime,l)\frac{d^{\mu_1\sigma_1}}{l^2}
\int\frac{d^4k_1}{(2\pi)^4}S^{\,b_1a_1a_2}_{\sigma_1\rho_1\rho_2}(p,\epsilon;l,k_1,k_2,p)\frac{d^{\rho_1\nu_1}d^{\rho_2\nu_2}}{k_1^2k_2^2}
S^{a_1a_2}_{A\,\nu_1\nu_2}(p_A,r_A,A^\prime,k_1,k_2)
\ee
where $S^{\,b_1}_{B\,\mu_1} = \langle p_{B^\prime},l|J_{\mu_1}^{b_1}(0)|p_b,r_B\rangle$
and $S^{a_1a_2}_{A\,\nu_1\nu_2}(p_A,r_A,A^\prime,k_1,k_2)$ give the amplitudes of finding gluon with color $b_1$, four-momentum $l$ and two gluons
of colors $a_1,a_2$ and momenta $k_1, k_2=k-k_1$ respectively in the decay products of a struck proton. The incoming proton polarizations are denoted by $r_{A,B}$. The vertex $S^{\,b_1a_1a_2}_{\sigma_1\rho_1\rho_2}(l,k_1,k_2; p,\epsilon)$ describes the amplitude of $J/\psi$ production with momentum $p$ and polarization $\epsilon$. In the standard Regge kinematics of high energy scattering at a small momentum transfer one can approximate polarization tensors of gluons (in the Feynman gauge) as $d^{\mu_1\sigma_1} \approx -2p_A^{\mu_1}p_B^{\sigma_1}/s$ and $d^{\rho_i\nu_i} \approx -2p_A^{\rho_i}p_B^{\nu_i}/s$ for $i=1,2$. Then, one can write the amplitude (\ref{amplitude1}) in the following form
\beq
-i\mathcal{M} = \frac{2s}{(2\pi)^4}%S^{\,b_1}_{B\,\mu_1}\frac{p_A^{\mu_1}}{s}
\Phi^{b_1} _{B}(p_B,r_B,B';\alpha_l, \beta_l, \nvec{l})\,
\frac{1}{l^2}
\int\frac{d^2\nvec{k}_{1}}{\nvec{k}_{1 }^{\,2}(\nvec{k} -\nvec{k}_{1 })^2}
\Phi^{a_1a_2}_A(p_A,r_A,A^\prime;\alpha_k,\beta_k;\nvec{k}_{1 },\nvec{k} -\nvec{k}_{1 }) \nonumber\\
\times \;\; \Phi^{b_1a_1a_2}_{J/\psi}(\alpha_k,\beta_l;\nvec{l} ,\nvec{k}_{1 },\nvec{k} -\nvec{k}_{1 };p,\epsilon),
\eeq
where the impact factors (at the amplitude level) for a $p_B \to B' + g$, $p_A \to A' + 2g$ and $2g + g \to J/\psi$ transitions read correspondingly:
\beq
\Phi^{b_1} _{B}(p_B,r_B,B';\alpha_l, \beta_l, \nvec{l}) & = & S^{\,b_1}_{B\,\mu_1}(p_B,r_B,B';l)\,\frac{p_A^{\mu_1}}{s},\\
\Phi^{a_1a_2}_A(p_A,r_A,A^\prime;\alpha_k,\beta_k;\nvec{k}_{1 },\nvec{k} -\nvec{k}_{1 }) &=&
\int d\beta_1^k S^{a_1a_2}_{A\,\nu_1\nu_2}(p_A,r_A,A^\prime,k_1,k-k_1)\frac{p_B^{\nu_1}p_B^{\nu2}}{s} ,\\
\Phi^{b_1a_1a_2}_{J/\psi}(\alpha_k,\beta_l;\nvec{l} ,\nvec{k}_{1 },\nvec{k} -\nvec{k}_{1 };p\,\epsilon) &=&
\int d\alpha_1^k S^{\,b_1a_1a_2}_{\sigma_1\rho_1\rho_2}(l,k_1,k-k_1;p,\epsilon)\frac{p_B^{\sigma_1}p_A^{\rho_1}p_A^{\rho_2}}{s} .
\eeq

\begin{figure}
\centerline{
a) \includegraphics[width=0.23\columnwidth]{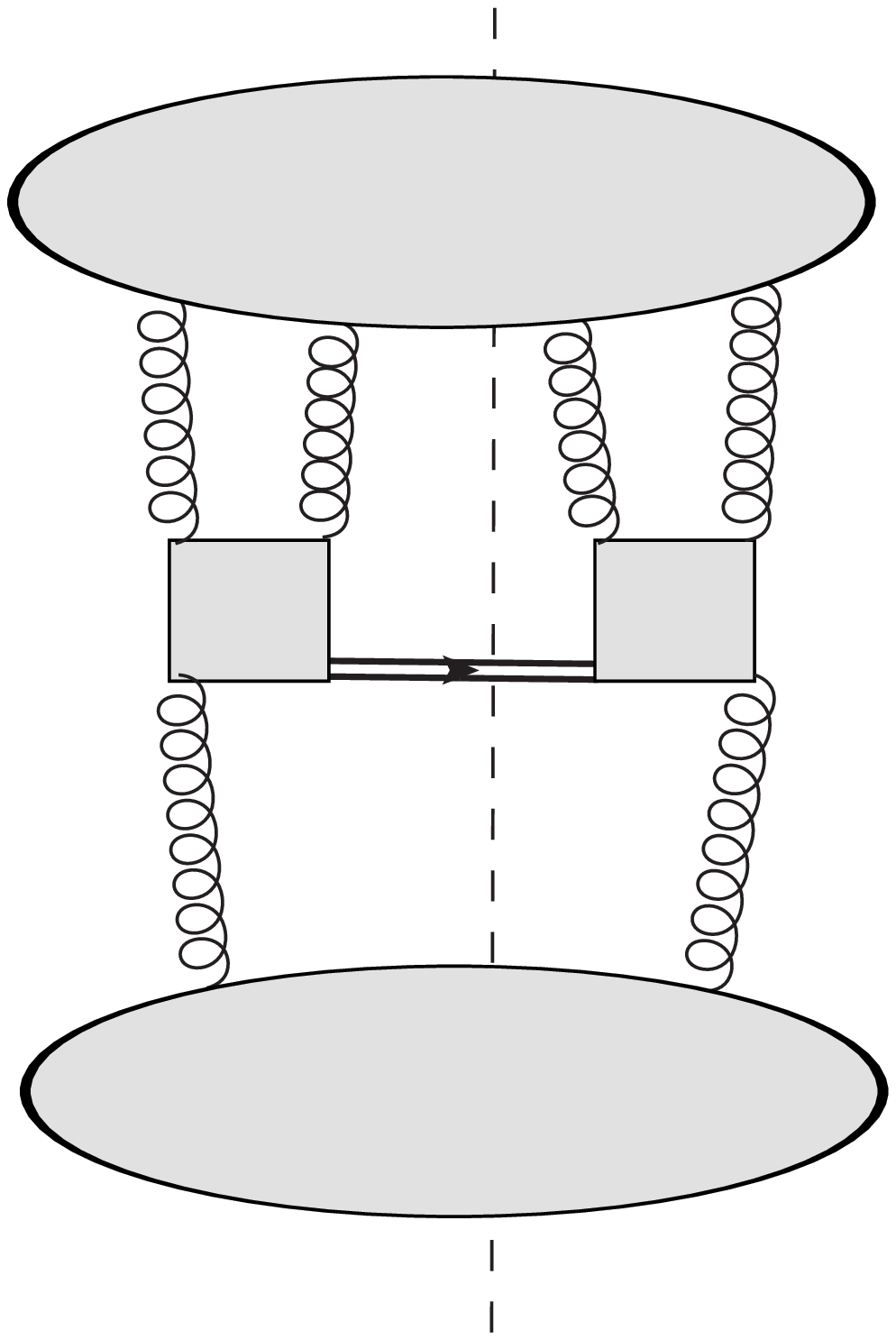}\hspace{10em}
b) \includegraphics[width=0.23\columnwidth]{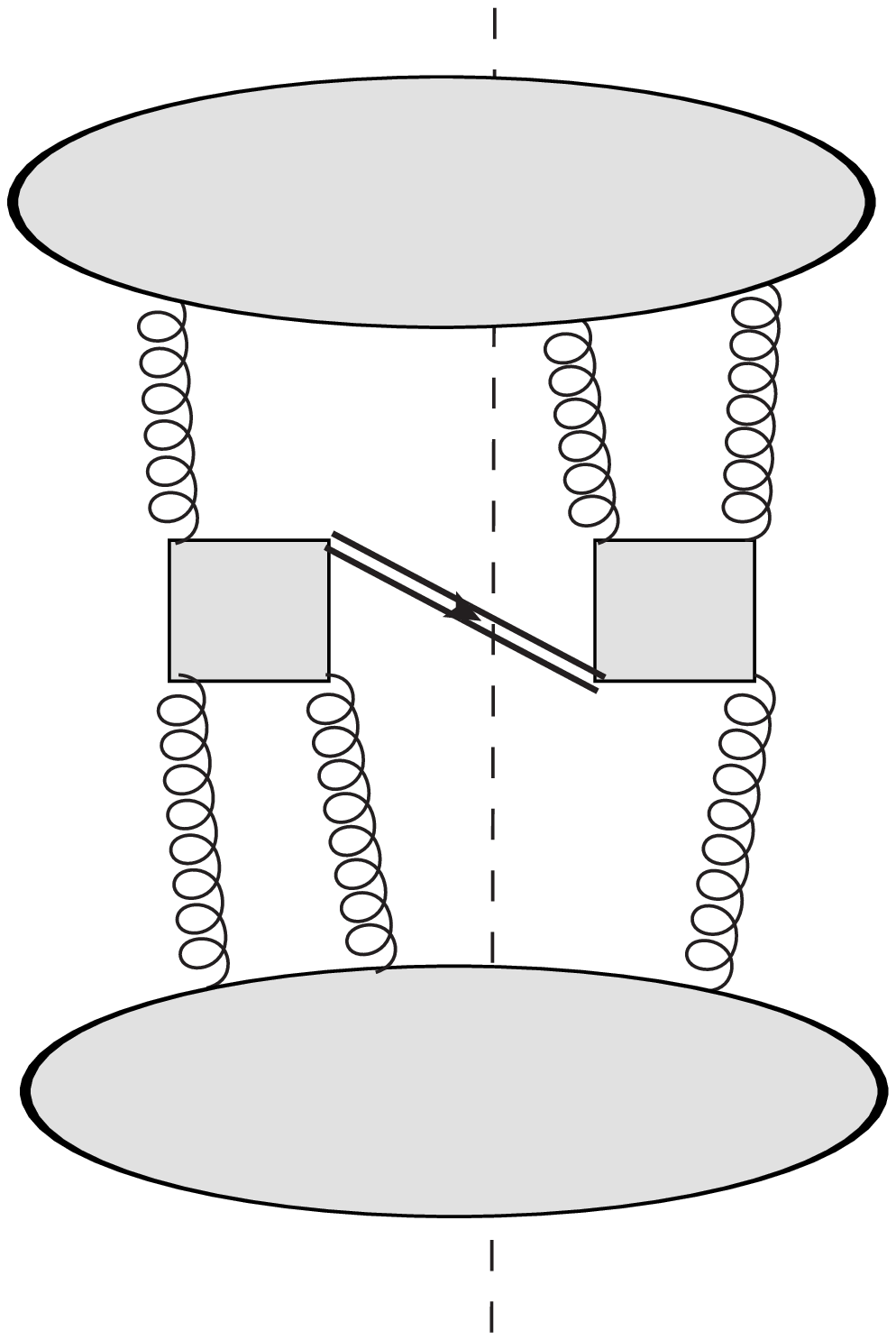}%\hspace{5em}
}
\caption{Topologies contributing to the rescattering correction in the $J/\psi$ cross-section : a) the topology with three two-gluon ladders (leading), b) the topology of two three-gluon BKP states (suppressed).
\label{fig2}
}
\end{figure}

The total contribution of three gluons to $J/\psi$ production amplitude comes from a diagram depicted in Fig.\ \ref{fig1}a and an `upside-down' diagram with the same topology, but with two gluons coupling to the lower vertex $p_B \to B'$. The contribution of the other diagram, $\mathcal{M}'$ may be obtained from $\mathcal{M}$, corresponding to the diagram in Fig.\ \ref{fig1}a, by exchange of kinematical variables: $\alpha \leftrightarrow \beta$. It is important to note that the interference of these two amplitudes, $\mathcal{M}$ and $\mathcal{M}'$ is a subleading effect. This is because terms $\mathcal{M}^*\mathcal{M}$ and $\mathcal{M}'^*\mathcal{M}'$ are driven by the QCD evolution of a four-gluon $t$-channel state (two ladders in the large $N_c$ limit) emerging from one proton and a two-gluon $t$-channel state from the other proton (see Fig.\ \ref{fig2}a) whereas the interference terms $\mathcal{M}^*\mathcal{M}'$ and $\mathcal{M}'^*\mathcal{M}$ are driven by subleading three-gluon $t$-channel states from both sides, see Fig.\ \ref{fig2}b. Hence, at large energies and for large scales, the interference terms, shown in Fig.\ \ref{fig2}b, may be neglected and the contribution to the $J/\psi$ production cross-section from three intermediate gluons $d\sigma \propto |\mathcal{M}|^2 + |\mathcal{M'}|^2$.
%For a similar reason, a possible interference between $2g + g \to J/\psi$ and $g + g \to J/\psi g$ should be subleading.

\subsection{The uncorrelated triple gluon cross-section}

The simplest model of the two-gluon distribution in the proton assumes lack of correlations of the gluons in the transverse plane. This means that the double gluon density is proportional to a product of independent single gluon distributions. Below we estimate the triple gluon contribution to $J/\psi$ hadroproduction in this scenario.

The contribution to $J/\psi$ cross-section, corresponding to the diagram shown in Fig.\ \ref{fig1}b takes the following form:
\beq
\label{cross-section1}
&&d\sigma_{pp\rightarrow J/\psi X} = \frac{2}{(2\pi)^8} \int\frac{d^4l}{(2\pi)^4}\frac{1}{(l^2)^2}
\Phi^{b_1b_2}_{2,p} (\alpha_l,\beta_l,\nvec{l}_{ }) \\\nonumber
&&\times \; \int\frac{d^4k}{(2\pi)^4}\int\frac{d^2\nvec{k}_1}{\nvec{k}_{1 }^{\,2}(\nvec{k}-\nvec{k}_{1 })^2}
\int\frac{d^2\nvec{k}_3}{\nvec{k}_{3 }^{\,2}(\nvec{k}-\nvec{k}_{3 })^2}
\Phi^{a_1a_2a_3a_4}_{4,p}(\alpha_k,\beta_k,\nvec{k}_{1 },\nvec{k}_{ }-\nvec{k}_{1 };\nvec{k}_{3 },\nvec{k}_{ }-\nvec{k}_{3 }) \\\nonumber
&& \times \; \Phi^{b_1a_1a_2}_{J/\psi}(\alpha_k,\beta_l;\nvec{l} ,\nvec{k}_{1 };,\nvec{k} -\nvec{k}_{1 },p\,;\epsilon)
   \Phi^{b_2a_3a_4\,\ast}_{J/\psi}(\alpha_k,\beta_l;\nvec{l} ,\nvec{k}_{3 },\nvec{k} -\nvec{k}_{3 },p\,;\epsilon) \\\nonumber
&& \times \; (2\pi)^4\delta(l+k-p)\frac{d^4p}{(2\pi)^4}\theta(p_0)2\pi\delta(p^2-M_{J/\psi}^2)\, + \,
 ( U \leftrightarrow L),
\eeq
where the last term describes contribution from the graph where upper and lower vertices from Fig.\ \ref{fig1}b are interchanged. The pomeron -- proton impact factor is defined by the formula
\be
\label{Phi2}
\Phi^{b_1b_2}_{2,p} (\alpha_l,\beta_l,\nvec{l}_{ }) = \frac{1}{2}\, \sum_{r_B}\sum_{B^\prime} \int d\Phi_{B^\prime}
(2\pi)^4\, \delta(l-p_B+p_{B^\prime})\, S^{\,b_1}_{B\,\mu_1}S^{\,b_2\,\ast}_{B\,\mu_2}\, \frac{p_A^{\mu_1}p_A^{\mu_2}}{s} \; ,
\ee
whereas
\beq
&&\Phi^{a_1a_2a_3a_4}_{4,p}(\alpha_k,\beta_k,\nvec{k}_{1 },\nvec{k}_{ }-\nvec{k}_{1 };\nvec{k}_{3 },\nvec{k}_{ }-\nvec{k}_{3 }) =
\frac{1}{2}\sum_{r_A}\sum_{A^\prime} \int d\Phi_{A^\prime} (2\pi)^4\delta(k-p_A+p_{A^\prime}) \\\nonumber
&&\times\; \Phi^{a_1a_2}_A(p_A,r_A,A^\prime;\alpha_k,\beta_k;\nvec{k}_{1 },\nvec{k} -\nvec{k}_{1 })
\Phi^{a_3a_4\,\ast}_A(p_A,r_A,A^\prime;\alpha_k,\beta_k,\nvec{k}_{3 },\nvec{k} -\nvec{k}_{3 })
\eeq
gives the two-pomeron--proton impact factor ($2\mathbb{P}$-p). In the last line the energy-momentum conservation delta function in the $J/\psi$ vertex gives $\alpha_k\approx \alpha, \beta_l\approx\beta$.

\begin{figure}
\centerline{
\includegraphics[width=0.5\columnwidth]{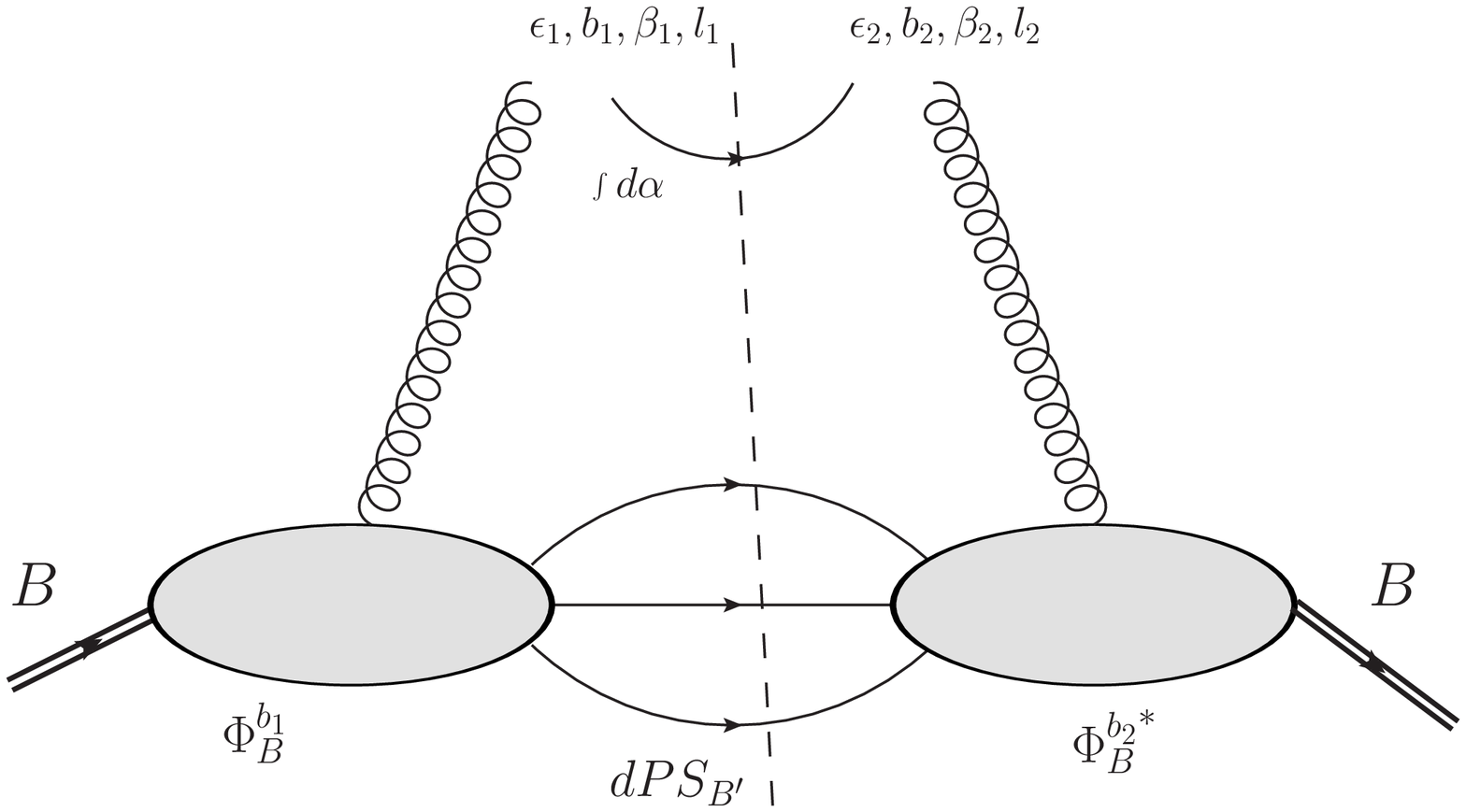}\hspace{5em}
}
\caption{
The unintegrated gluon distribution.
\label{fig3}
}
\end{figure}

Several remarks are in order here:
\begin{enumerate}
\item The pomeron -- proton impact factor $\Phi_{2,p}$ (see Fig.\ \ref{fig3}) integrated over longitudinal variables gives
\be
\int d\alpha_l d\beta_l\Phi^{b_1b_2}_{2,p} (\alpha_l,\beta_l,\nvec{l}_{ }) = \frac{(2\pi)^3}{N_c^2-1} f(\beta,l^2)\delta^{b_1b_2},
\ee
where $\beta_l\approx \beta$ follows from the energy-momentum conservation in $J/\psi$ vertex and $f(\beta,l^2)$ is an unintegrated gluon distribution.
\item The $2\mathbb{P}$--p impact factor integrated over longitudinal variable $\beta_k$ can be decompose in the following way(see Appendix for the derivation):
\beq
\label{Phi4}
&&\int d\alpha_k d\beta_k
\Phi^{a_1a_2a_3a_4}_{4,p}(\alpha_k,\beta_k,\nvec{k}_{1 },\nvec{k}_{ }-\nvec{k}_{1 };\nvec{k}_{3 },\nvec{k}_{ }-\nvec{k}_{3 }) = \nonumber\\
&&\frac{R_{\mbox{sh}}^2(2\pi)^7}{(N_c^2-1)^2s}
\left[\delta^{a_1a_3}\delta^{a_2a_4}
f(\alpha,k_1 ^2)f(\alpha,(\nvec{k}_{ }-\nvec{k}_{1 })^2)
S^2(\nvec{k}_{1 }-\nvec{k}_{3 }) \right. \nonumber \\
&&\left. + \; \delta^{a_1a_4}\delta^{a_2a_3}
f(\alpha,k_1^2)f(\alpha,(\nvec{k}_{ }-\nvec{k}_{1 })^2)
S^2(\nvec{k}_{1 }+\nvec{k}_{3 }-\nvec{k} )
+  \delta^{a_1a_2}\delta^{a_3a_4}...\right]\, ,
\eeq
where $\alpha_k\approx \alpha$ follows from the energy-momentum conservation in $J/\psi$ vertex. The last term in parenthesis does not contribute to the cross-section due to the color factor contraction with the $J/\psi$ vertex. $S(\nvec{k})$ is a symmetric function peaked around $\nvec{k}=0$ and $R_{\mathrm{sh}}$ is the Shuvaev factor \cite{Shuvaev} introduced to account for the fact that longitudinal variables of the gluons are not equal and the off-diagonal gluon distributions enter the cross-section.

\item The $J/\psi$ triple gluon vertex reads \cite{BMSC}:
\beq
\label{PhiJPsi}
\Phi^{b_1a_1a_2}_{J/\psi}(\alpha,\beta;\nvec{l} ,\nvec{k}_{1 };,\nvec{k} -\nvec{k}_{1 },p\,;\epsilon) &=& g^3\frac{d^{a_1a_2b_1}}{N_c}
V_{J/\psi}(\alpha,\beta;\nvec{l}_{ },\nvec{k}_{1 },\nvec{k}_{2 };\epsilon),\\\nonumber
V_{J/\psi}(\alpha,\beta;\nvec{l}_{ },\nvec{k}_{1 },\nvec{k}_{2 };\epsilon) &=& 4\pi m_c g_{J/\psi}
\left[-\frac{\epsilon^\ast\cdot (x p_A + l_{\perp})}{\nvec{l}_{ }^{\,2}+(\nvec{k}_{1 }+\nvec{k}_{2 })^2+4m_c^2} \right.  \\\nonumber
&+& \left. \frac{\epsilon^\ast\cdot p_A (x-\frac{4\nvec{k}_{1 }\cdot\nvec{k}_{2 }}{y s})+\epsilon^\ast\cdot l_{\perp}}
{\nvec{l}_{ }^{\,2}+(\nvec{k}_{1 }-\nvec{k}_{2 })^2+4m_c^2}\right],
\eeq
where $\nvec{k}_2=\nvec{k}-\nvec{k}_1$.
\end{enumerate}

Substituting equations (\ref{Phi2}) through (\ref{PhiJPsi}) into (\ref{cross-section1}) performing delta function integrations and color factor contractions one arrives at the final formula for the rescattering correction to polarization dependent differential $J/\psi$ hadroproduction cross-section,
%\beq
%\label{cross-section2}
%&& \frac{d^3\sigma_{pp\rightarrow J/\psi X}}{d\ln\beta dp^2_\perp} = \frac{2\alpha_s^3}{\pi}\frac{N_c^2-4}{N_c^3(N_c-1)^2}R_{\mbox{sh}}^2
%\int d^2\nvec{k}d^2\nvec{k}_{1}
%\frac{f(\beta,\nvec{p}_{ }-\nvec{k}_{ })f(\alpha,\nvec{k}_{1 })f(\alpha,\nvec{k}_{ }-\nvec{k}_{1 })}
%{\left((\nvec{p} -\nvec{k}_{ })^2\nvec{k}_{1 }^{\,2}(\nvec{k}-\nvec{k}_{1 })^2\right)^2} \nonumber\\
%&&\left|V_{J/\psi}(\alpha,\beta;\nvec{k}_{1 },\nvec{k}-\nvec{k}_{1 },\nvec{p}_{ }-\nvec{k}_{ };\epsilon)\right|^2
%\int d^2\nvec{k}^{\prime\prime}_\perp S^2(\nvec{k}^{\prime\prime}_\perp) + (\alpha\leftrightarrow\beta,p_A\leftrightarrow p_B) .
%\eeq
\beq
\label{cross-section2}
&& \frac{d^3\sigma_{pp\rightarrow J/\psi X}}{d\ln\beta dp^2_\perp} = \frac{2}{\pi}\frac{N_c^2-4}{N_c^3(N_c-1)^2}\frac{R_{\mbox{sh}}^2}{\sigma_{\mathrm{eff}}}
\int d^2\nvec{k}d^2\nvec{k}_{1}
\frac{\alpha_s^3 f(\beta,(\nvec{p}_{ }-\nvec{k}_{ })^2)
f(\alpha,k_1 ^2)
f(\alpha,(\nvec{k}_{ }-\nvec{k}_{1 })^2)}
{\left((\nvec{p} -\nvec{k}_{ })^2\nvec{k}_{1 }^{\,2}(\nvec{k}-\nvec{k}_{1 })^2\right)^2} \nonumber\\
&&\times \; \left|V_{J/\psi}(\alpha,\beta;\nvec{k}_{1 },\nvec{k}-\nvec{k}_{1 },\nvec{p}_{ }-\nvec{k}_{ };\epsilon)\right|^2
 + (\alpha\leftrightarrow\beta,p_A\leftrightarrow p_B) ,
\eeq
where the last term describes contribution from the graph as in Fig.\ \ref{fig1}b where four gluons are attached to the $p_B$ instead of $p_A$ vertex. In the derivation one needs to evaluate an integral $\int d^2\nvec{k}_\perp S^2(\nvec{k}_\perp)$ which, assuming the Gaussian form of $\tilde S(\nvec{b})$, may be related to the inverse of multiple scattering parameter, $\int d^2\nvec{k}_\perp S^2(\nvec{k}_\perp)  = 8\pi^2 \sigma^{-1}_{\mathrm{eff}}$.

\subsection{Triple gluon cross-section with gluon correlations}

Formula (\ref{cross-section2}) describes the distributions of transverse positions of two-gluons coming from one of the protons as being independent, i.e.\ uncorrelated, in the transverse plane. At large $p_T$ of the meson one should, however, expect some correlations of these distribution to emerge in the course of QCD evolution of double gluon density. In this evolution \cite{BoFKL}, a single parton ladder may split into two ladders, emerging at the same impact parameter, that introduces correlations of the parton positions.

\begin{figure}
\centerline{
\includegraphics[width=0.4\columnwidth]{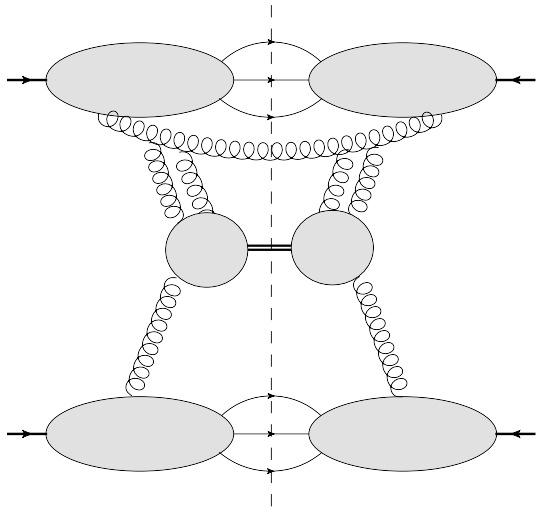}
}
\caption{The correlated triple gluon  contribution.
\label{fig4}}
\end{figure}

A diagram that describes the gluon parton splitting into two gluons is shown in Fig.\ \ref{fig4}. This contribution to $J/\psi$ hadroproduction was discussed in \cite{KMRS} in the collinear limit and using only an approximate estimate of QCD evolution of the double gluon state. This correlated rescattering correction is potentially important as it is expected to lead to a less steep $p_T$ dependence of the resulting differential cross-section component than the one obtained with the uncorrelated gluon distributions. We calculated this amplitude with full $k_T$ dependence of the three gluons that enter the $J/\psi$ vertex, an assuming that the initial gluon in a upper vertex, that acts as a source of the double gluon, is collinear with the parent proton. After standard steps we obtained the following cross-section,
%\beq
%\frac{d^3\sigma_{pp\rightarrow J/\psi X}}{d\ln\beta dp^2_\perp} &=& \frac{\alpha_s^3\alpha_s^2}{\pi^4}\frac{9}{4}\frac{N_c^2-4}{N_c^3(N_c-1)}
%\int g(\xi)d\xi
%\int\frac{d^2\nvec{k}d^2\nvec{k}_{1}d^2\nvec{k}_{2}f(\beta,\nvec{p}_{ }-\nvec{k}_{ })}
%{(\nvec{p} -\nvec{k}_{ })^4\nvec{k}_{1 }^{\,2}(\nvec{k}-\nvec{k}_{1 })^2\nvec{k}_{2 }^{\,2}(\nvec{k}-\nvec{k}_{2 })^{\,2}} \nonumber\\
%&& V_{J/\psi}(\alpha,\beta;\nvec{k}_{1 },\nvec{k}-\nvec{k}_{1 },\nvec{p}_{ }-\nvec{k}_{ };\epsilon)
%V^\ast_{J/\psi}(\alpha,\beta;\nvec{k}_{2 },\nvec{k}-\nvec{k}_{2 },\nvec{p}_{ }-\nvec%{k}_{ };\epsilon)
%\eeq
\beq
\label{eq:sig_correlated}
\frac{d^3\sigma_{pp\rightarrow J/\psi X}}{d\ln\beta dp^2_\perp} &=& \frac{\alpha_s^3\alpha_s^2}{\pi^4}\frac{9}{4}\frac{N_c^2-4}{N_c^3(N_c-1)}
\int g(\xi,\mu)d\xi
\int\frac{d^2\nvec{k}d^2\nvec{k}_{1}d^2\nvec{k}_{2}
f(\beta,(\nvec{p}_{ }-\nvec{k}_{ })^2)}
{(\nvec{p} -\nvec{k}_{ })^4\nvec{k}_{1 }^{\,2}(\nvec{k}-\nvec{k}_{1 })^2\nvec{k}_{2 }^{\,2}(\nvec{k}-\nvec{k}_{2 })^{\,2}} \nonumber\\
&& \times \; V_{J/\psi}(\alpha,\beta;\nvec{k}_{1 },\nvec{k}-\nvec{k}_{1 },\nvec{p}_{ }-\nvec{k}_{ };\epsilon)\,
V^\ast _{J/\psi}(\alpha,\beta;\nvec{k}_{2 },\nvec{k}-\nvec{k}_{2 },\nvec{p}_{ }-\nvec{k}_{ };\epsilon).
\eeq
An analogous formula is valid also for a quark from the target replacing the gluon as a source of the two correlated $t$-channel gluon after the suitable replacement of the pdfs and the color factor. This formula provides the lowest order approximation to the correlated rescattering amplitude, which is only an input amplitude for the evolution of the two-gluon state from the parent gluon to the $J/\psi$ vertex. Since the typical evolution length in rapidity is sizable for this evolution one expects a potentially large effects of the QCD evolution to occur. The treatment of this evolution in the framework of the small~$x$ resummation is possible but it is a highly nontrivial task and it is beyond the scope of the present analysis. However, we estimated numerically the lowest order contribution of (\ref{eq:sig_correlated})  and found it to be significantly smaller than the uncorrelated rescattering contribution of (\ref{cross-section2}). Thus in the present analysis this contribution will be disregarded.

\section{Numerical results}
\label{sec:4}

\subsection{Triple gluon corrections in $p\bar p$ and $pp$ collisions.}

\begin{figure}
\centerline{
\begin{tabular}{ll}
a) \hspace{1em}\includegraphics[width=0.47\columnwidth]{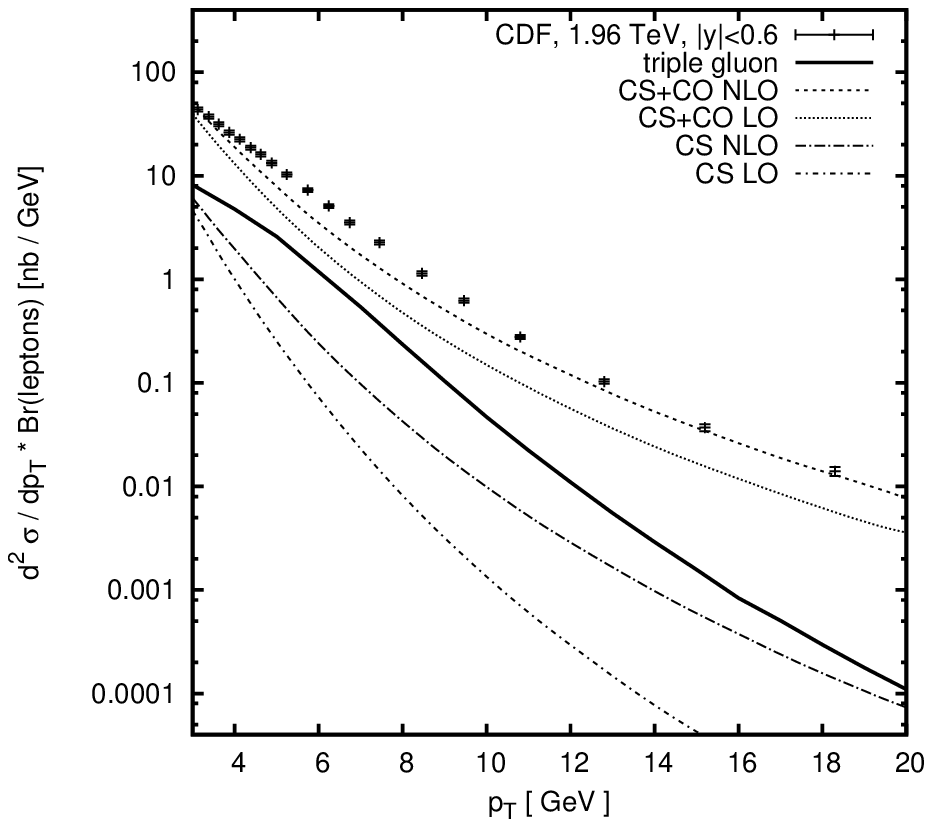}\hspace{1em} &
b) \hspace{1em}\includegraphics[width=0.47\columnwidth]{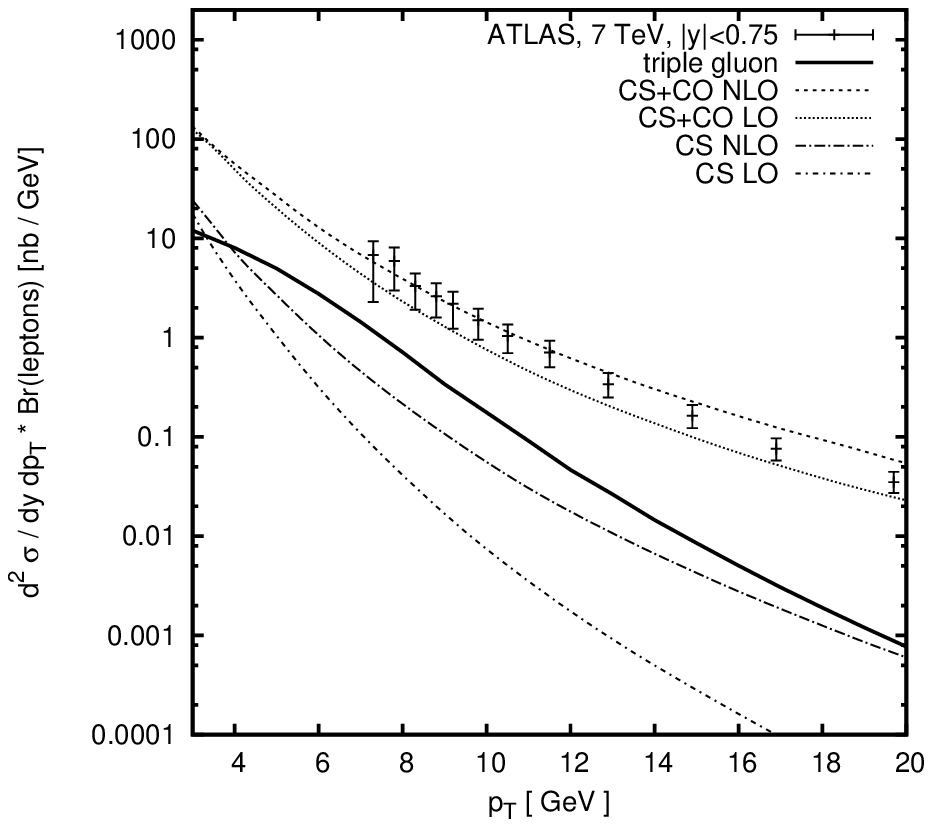}
\end{tabular}
}
\caption{Differential cross-sections for $J/\psi$ hadroproduction:
a) $\left. \frac{d\sigma_{pp\rightarrow J/\psi X}}{dp_T}\right|_{|Y|<0.6} \times \mathrm{Br}(J/\psi \to \mu^+ \mu^-)$ at the Tevatron, $\sqrt{s} = 1.96$~TeV and
b) $\left. \frac{d\sigma_{pp\rightarrow J/\psi X}}{dY\, dp_T}\right|_{|Y|<0.75} \times \mathrm{Br}(J/\psi \to \mu^+ \mu^-)$ at the LHC, $\sqrt{s} = 7$~TeV.
The data-points shown: a) CDF \cite{CDFdata} and b) ATLAS \cite{ATLASdata}. The curves shown: bold continuous --- the triple gluon contribution computed in this paper, the dashed and dotted thin curves represent results of Ref.~\cite{BK-NLOfit} in the collinear approach: color singlet (LO and NLO) and combined color octet and color singlet results (see the legend in the plots).
\label{fig5}
}
\end{figure}

In the numerical evaluations the unintegrated gluon densities were used derived from the CT10 collinear gluon density \cite{ct10} using Kimber-Martin-Ryskin (KMR) scheme~\cite{KMR} with the factorization scale given by the transverse meson mass, $\mu^2 = M_{J/\psi} ^2 + p_T^2$. The running strong coupling constant of a gluon with virtuality $k^2$ was evaluated at the scale $\mu_0 ^2 = M_c^2 + k^2$, with $M_c = M_{J/\psi} /2$. This scale choice is consistent with the KMR prescription. We set the multiple scattering parameter value $\sigma_{\mathrm{eff}} = 15$~mb, in accordance with the experimental results from the Tevatron \cite{sigma0cdf,sigma0d0}. For the effective gluon distribution $\lambda$ in the Shuvaev factor $R_{\mathrm{Sh}}$ we set $\lambda = 0.3$, which leads to $R_{\mathrm{Sh}} = 1.3$.

In order to visualise the relevance of the triple gluon correction we compare the obtained results to a limited choice of experimental results from the Tevatron and the LHC, and to the standard collinear QCD fits within the color singlet and octet models. In this paper we do not aim to fit the data, nor provide the global description,  so for clarity, only $p_T$ dependencies near the central rapidity $Y=0$ are shown.   

In Fig.\ \ref{fig5}a the results of numerical evaluation of the triple gluon correction to the differential meson production cross section  $\left. \frac{d\sigma_{pp\rightarrow J/\psi X}}{dp_T}\right|_{|Y|<0.6} \times \mathrm{Br}(J/\psi \to \mu^+ \mu^-)$ (including the branching ratio of the meson decay to muons) are shown for the Tevatron energy ($\sqrt{s} = 1.96$ GeV) --- the bold continuous line. For reference we display also the CDF data \cite{CDFdata} and the results of collinear calculations of the color singlet and CS+CO predictions at LO and NLO \cite{BK-NLOfit}. Clearly, the triple gluon contribution enters as a subleading correction to the prompt $J/\psi$ cross-section that may reach 20-25\% of the total cross-section at moderate $p_T$ and becomes negligible at larger $p_T$. On the other hand, the triple gluon contribution exceeds the standard (collinear) color singlet contribution in the relevant $p_T$ range.

In Fig.\ \ref{fig5}b we also show a similar comparison for the ATLAS data in the central rapidity region. Note that the data-points describe here the double-differential cross-section, $\left. \frac{d\sigma_{pp\rightarrow J/\psi X}}{dYdp_T}\right|_{|Y|<0.75} \times \mathrm{Br}(J/\psi \to \mu^+ \mu^-)$. The overall pattern of different contributions and the data-points is similar to the pattern described above for the Tevatron energies. Again, the triple gluon contribution exceeds the CSM contributions, but makes not more than  20-25\%  of the total prompt $J/\psi$ cross-section, and the triple gluon correction becomes negligible at larger $p_T$.

\begin{figure}
\begin{center}
\begin{tabular}{lll}
\includegraphics[width=0.34\columnwidth]{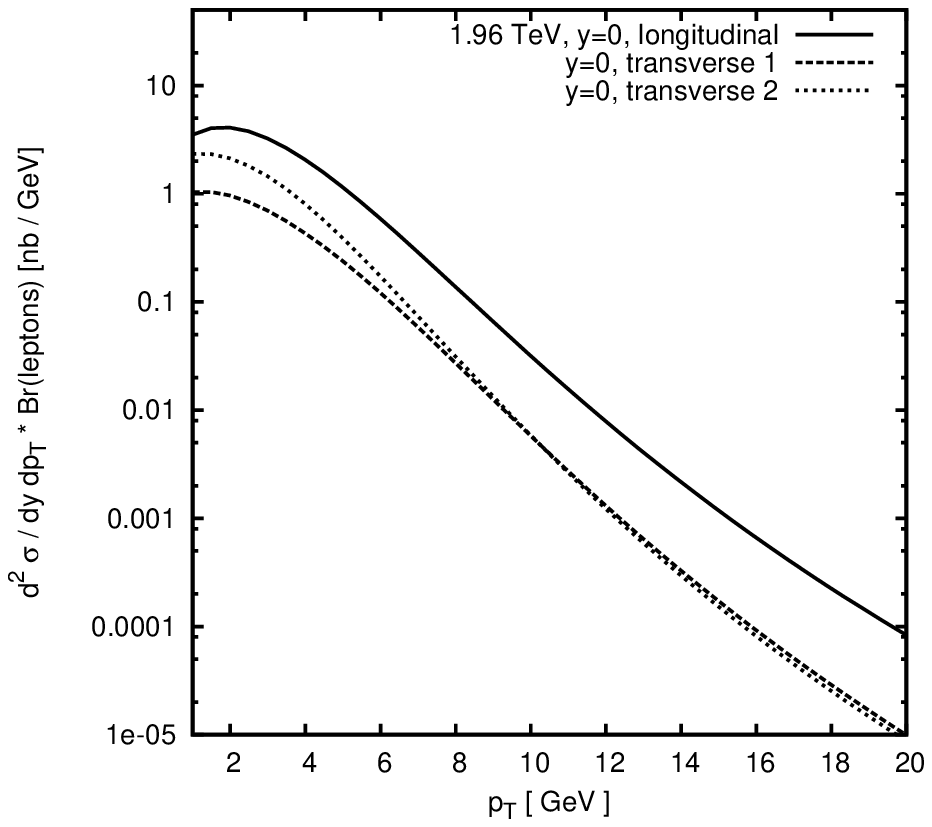}\hspace{-2em}&
\includegraphics[width=0.34\columnwidth]{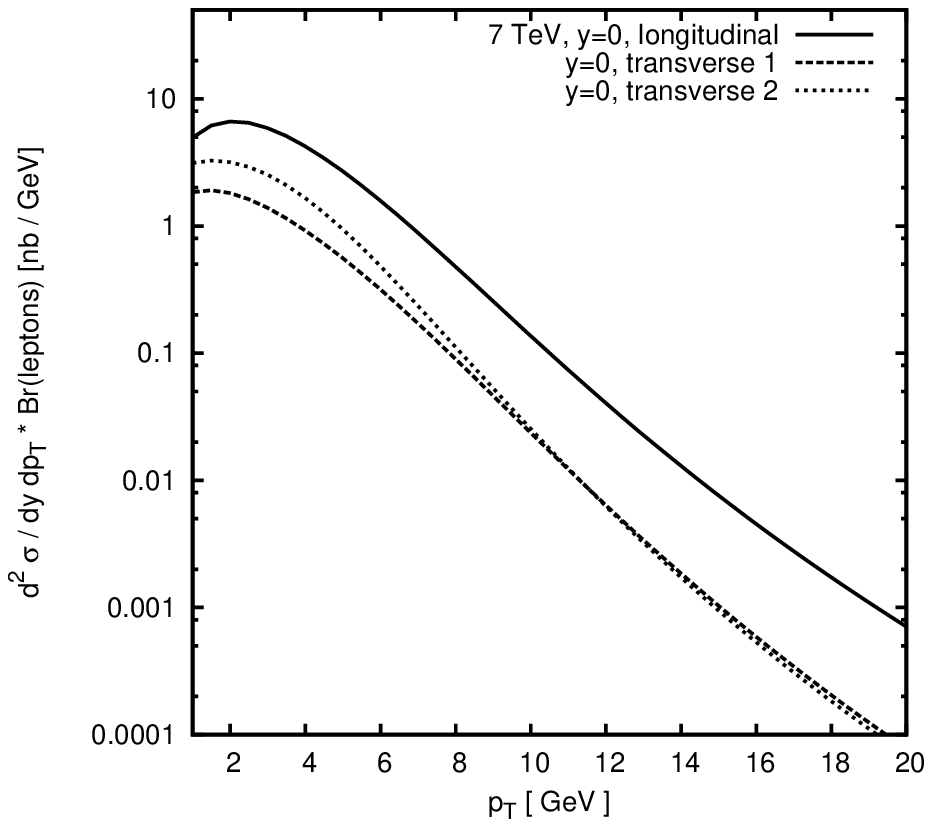}\hspace{-2em}&
\includegraphics[width=0.34\columnwidth]{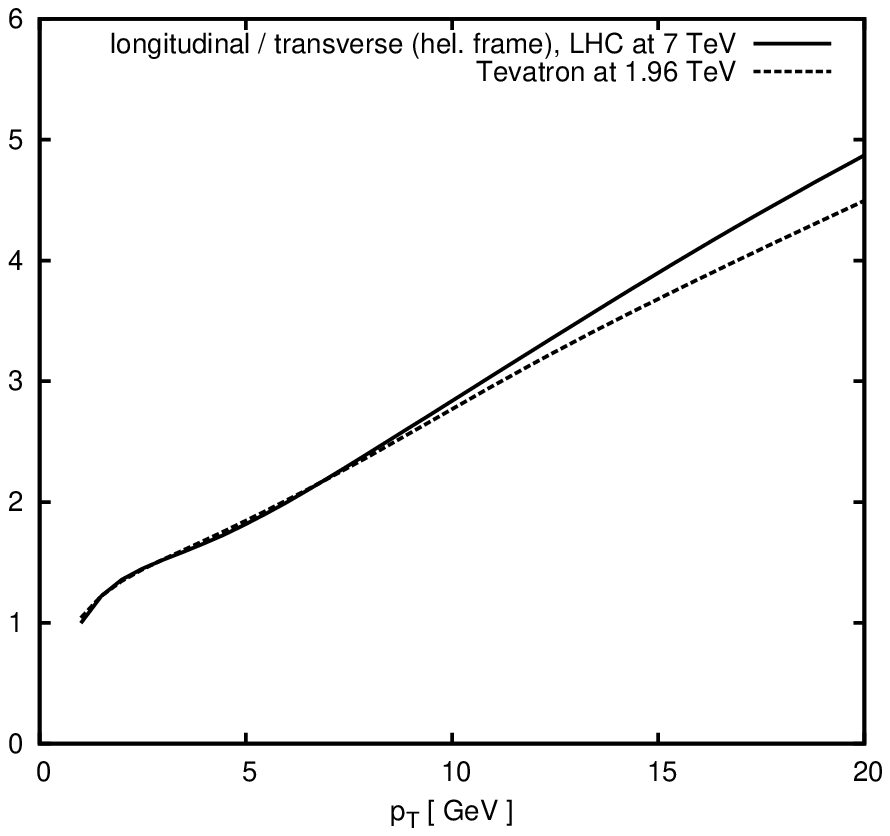}\hspace{-2em}\\
a) & b) & c)
\end{tabular}
\end{center}
\caption{Polarization components of the triple gluon correction:
a) $\left. d^2 \sigma_{p\bar p\to J/\psi(\sigma) X} / dY dp_T\right| _{Y=0}$ for the Tevatron at $\sqrt{s} = 1.96$~TeV, b) $\left. d^2 \sigma_{pp\to J/\psi(\sigma) X} / dY dp_T\right| _{Y=0}$ for the LHC at $\sqrt{s} = 7$~TeV, c) ratio of the cross-section for
longitudinal and transverse polarizations. In a) and b) the polarizations $\sigma$ are chosen in the helicity frame. }
\label{fig6}
\end{figure}

In Fig.\ \ref{fig6}a polarized components are shown of the triple gluon correction in the helicity frame for the Tevatron (at $\sqrt{s} = 1.96$~TeV) and in Fig.\ \ref{fig6}b for the LHC (at $\sqrt{s} = 7$~TeV) and for the central rapidity, $Y=0$. The transverse polarization~1 in the plot is referred to the transverse polarization contained in the plane spanned by the beam axis and the meson three-momentum in the laboratory frame, and the transverse polarization~2 is perpendicular to this plane. Clearly, at low transverse momentum the longitudinal and the total transverse cross-sections are close to each other, and with increasing $p_T$ the longitudinal component becomes dominant, almost saturating the total triple gluon correction at $p_T = 20$ ~GeV. As seen from Fig. \ref{fig6}c the pattern of polarized cross-section for the Tevatron and the LHC is very similar.

The presented results for the triple gluon correction in $J/\psi$ hadroproduction are our central theoretical predictions. There is, however, theoretical uncertainty of these results coming from scale choices in the running coupling constant, the details of the unintegrated gluon densities, and from the unknown higher order corrections. The full analysis of these uncertainties is beyond the scope of this paper, however, we performed some first estimates and tested sensitivity of the results to the $\mu_0$ scale variation in $\alpha_s (\mu_0)$ and collinear parton choice in the KMR scheme. The results depend weakly on the collinear parton set, however, the increase of $\mu_0$ to $2\mu_0$ leads to reduction of the cross-sections by about 50\% in the moderate $p_T$ range. For increasing $p_T$ this effect is getting weaker. On the other hand, applications of the unintegrated gluon densities (and related schemes) CCFMA0 \cite{ccfma0} and JH2013 \cite{JH2013} instead of the KMR scheme gave results larger by a factor of 2--3 in moderate $p_T$-s (say for $p_T < 5$~GeV), and at larger $p_T$-s (say for $p_T>7$~GeV) the CCFMA0 and JH2013 results are close to our central predictions. Summarizing these theoretical uncertainty checks, the normalization of our predictions in uncertain by a factor~2 up or down for moderate $p_T$-s, and the normalization and shape of the distributions at large $p_T$ is much less uncertain.

\subsection{Probing the triple gluon contribution with nuclear beams}

In $pp$ and $p\bar p$ the contributions of triple (or multiple) gluon exchange mechanisms of heavy vector meson hadroproduction may be not easy to disentangle from the standard mechanism of production within the CSM and COM. In particular, the triple gluon contribution may be to some extent absorbed into fit parameters of the COM. A more sensitive probe of the triple and multiple gluon exchange effects in vector quarkonia hadroproduction should be provided in experiments with nuclear beams, in particular, in proton-nucleus collisions, $pA$, where $A$ is the number of nucleons in the nucleus. In such collisions the triple gluon contribution scales with $A$ in different way than the standard (two-gluon) contributions.

Let us write the $pA$ cross-section $\sigma^{pA}_V$ as a sum of the standard gluon-gluon contribution $\sigma^{pA}_{gg}$,  a contribution with two gluons coming from the proton $\sigma^{pA} _{(gg)g}$, and a contribution with two gluon coming from a nucleus $\sigma^{pA} _{g(gg)}$,
\be
\sigma^{pA} _V = \sigma^{pA}_{gg} + \sigma^{pA} _{(gg)g} + \sigma^{pA} _{g(gg)}.
\ee
In particular, for the $pp$ scattering it reduces to $\sigma^{pp} _V = \sigma^{pp}_{gg} + \sigma^{pp} _{(gg)g} + \sigma^{pp} _{g(gg)}$. Since in $\sigma^{pA}_{gg}$  and in $\sigma^{pA} _{(gg)g}$ one probes a single density of gluons in the nucleus, these cross-sections are enhanced by a factor of $A$ w.r.t.\ the proton-proton cross-sections,  $\sigma^{pA} _{gg} = A\sigma^{pp} _{gg}$ and $\sigma^{pA} _{(gg)g} = A\sigma^{pp} _{(gg)g}$.

The triple gluon contribution to $J/\psi$ production with two gluons coming from the nucleus is enhanced by $A^2 R^2 _p/ R^2_A$, with $R_p$ and $R_A$ being the proton and the nucleus effective radius. This dependence on the radius $R_A$ comes from the integral of the nuclear impact parameter density $S_A(\nvec{b})$ of the gluon distribution, $\int d^2 \nvec{b} S_A^2(\nvec{b}) \sim 1/ R_A^2$. Since $R_A \sim A^{1/3}R_p$,  we have
$\sigma^{pA} _{g(gg)} = A^{4/3}\sigma^{pp} _{g(gg)}$. Combining the $A$-dependencies of the chosen cross-section components one finds for the nuclear modification factor,
%per-nucleon cross-section ratio,
\be
\bar R_{pA} = \frac{\sigma^{pA} _V}{A\sigma^{pp} _V} = 1
+ (A^{1/3} -1)\frac{\sigma^{pp} _{g(gg)}}{\sigma^{pp}_V}.
\ee
Thus the deviation of $\bar R_{pA}$ from one, $\delta_A = \bar R_{pA} - 1$, measures the nuclear effects of multiple gluons from the nucleus. In our approach we neglect the effects of more than two gluons coming from the nucleus which is only motivated if the triple gluon correction is not large, $\delta_A \ll 1$.

Based on the obtained results on triple gluon contribution we also performed an estimate of $\delta_A$ for ALICE experiment conditions, that is for $J/\psi$ inclusive production in $p$Pb collisions at $\sqrt{s} = 5.02$~TeV, $-4.46 < Y < -2.96$ (the nucleus fragmentation region) and $2.03 < Y < 3.53$ (the forward region). We found $\delta_A \simeq 20\%$ for in the backward region and $\delta_A \simeq 100\%$ in the forward region. The result for the backward region is slightly larger from the ALICE measurements showing evidence for about 10\% nuclear enhancement of $R_{pA}$ with experimental errors of about 10\% \cite{ALICEdata}. Thus our estimates of the triple gluon correction are at the upper limit of ALICE measurements and more precise measurements may be used to constrain the details of our calculation and of the unintegrated gluon densities. The available data prefer a conservative approach of our central theory prediction with the unintegrated gluon densities derived from the collinear gluon densities within the KMR scheme.

In the forward region the triple gluon correction to $\bar R_{pA}$ is found to be large, over $100\%$. This implies that higher order rescattering corrections, with more than two gluons from the nucleus, are also important and the triple gluon contribution $\sigma^{pA} _{g(gg)}$ does not provide a reliable estimate of nuclear effects in the forward region for $p$Pb collisions at the LHC. 
%Also, one should in this region shadowing corrections to the standard processes are expected to be important. 
Therefore, in this region one should perform resummation of multiple scattering effects, as it was done e.g.\ in the color glass condensate approach~\cite{MV,KMV}. We conclude that the triple gluon contributions to $J/\psi$ hadroproduction may be experimentally constrained in $pA$ measurements with lighter nuclei and/or  central to backward meson rapidities where the triple gluon contribution is expected to bring a dominant correction to $\bar R_{pA}$. The key observable for such measurement is the $A$-dependence of the nuclear modification factor $\bar R_{pA}$.

\section{Conclusions}
\label{sec:5}

The performed calculations show that the uncorrelated triple gluon contribution to $J/\psi$ hadroproduction introduce a sizable $\sim 20$\% correction to theoretical predictions of prompt $J/\psi$ hadroproduction. The relative correction is largest for moderate transverse momenta $p_T < 8$~GeV and it quickly becomes negligible with $p_T$ growing beyond this range. The result for moderate $p_T$ is uncertain by about factor of~$2$ (up or down), but the large $p_T$ shape and normalization is more stable against variations of the model details. It follows, that the uncorrelated color singlet triple gluon contribution is unable to explain the large excess of the $J/\psi$ hadroproduction cross-section over the collinear color singlet predictions, especially at larger values of $p_T$. The triple gluon contribution, however, is found to be larger than the NLO color singlet cross-sections and therefore it should be relevant for theoretical analyses of heavy vector quarkonia hadroproduction. In particular, the estimated triple gluon contribution exhibits strong $p_T$ dependence of the meson polarization. At small $p_T$ this mechanism leads to equal rates of longitudinally and transversally polarized mesons, whereas at larger $p_T$ the longitudinal polarization strongly dominates. So the rescattering correction may modify the polarization pattern following from the collinear COM or KTF descriptions and it should be advantageous to complement with this contribution the fits to quarkonia hadroproduction cross-sections.

For proper theoretical understanding of heavy vector meson hadroproduction it should be important to constrain experimentally the triple gluon (or rescattering) corrections. This may be not easy in $pp$ and $p\bar p$ collisions where free parameters of the collinear COM or KTF fits may be used to absorb this correction. The multiple gluon effects, however, are characterized by a distinct (nonlinear) dependence of the gluon densities of colliding beams. Such effects may be directly probed in proton-nucleus ($pA$) collisions, where the relative triple gluon correction should be enhanced by a factor of $A^{1/3}$. Recent ALICE data on inclusive $J/\psi$ production in $p$Pb collisions at $\sqrt{s} = 5$~TeV in the nucleus fragmentation region show some evidence of such non-linear nuclear enhancement, however, more extended and accurate measurements are needed to determine experimentally the magnitude of the triple gluon correction with satisfactory precision.

\section*{Acknowledgements}
We are grateful to Hannes Jung, Sergey Baranov and Nikolai Zotov for interesting discussions and sharing computer codes. Support of the Polish National Science Centre grant no.\ DEC-2011/01/B/ST2/03643 is gratefully acknowledged. \\

\appendix
\section{}

\begin{figure}[h]
\centerline{
\includegraphics[width=0.45\columnwidth]{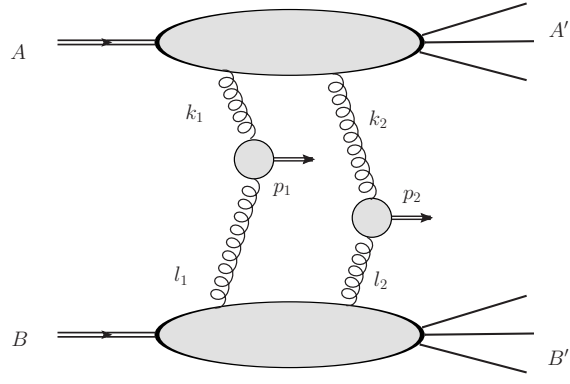}%\hspace{5em}
}
\caption{The amplitude for a semi-inclusive two particle production process $pp\rightarrow V_1V_2X$ \label{fig7}
}
\end{figure}

\begin{figure}
\centerline{
a) \includegraphics[width=0.4\columnwidth]{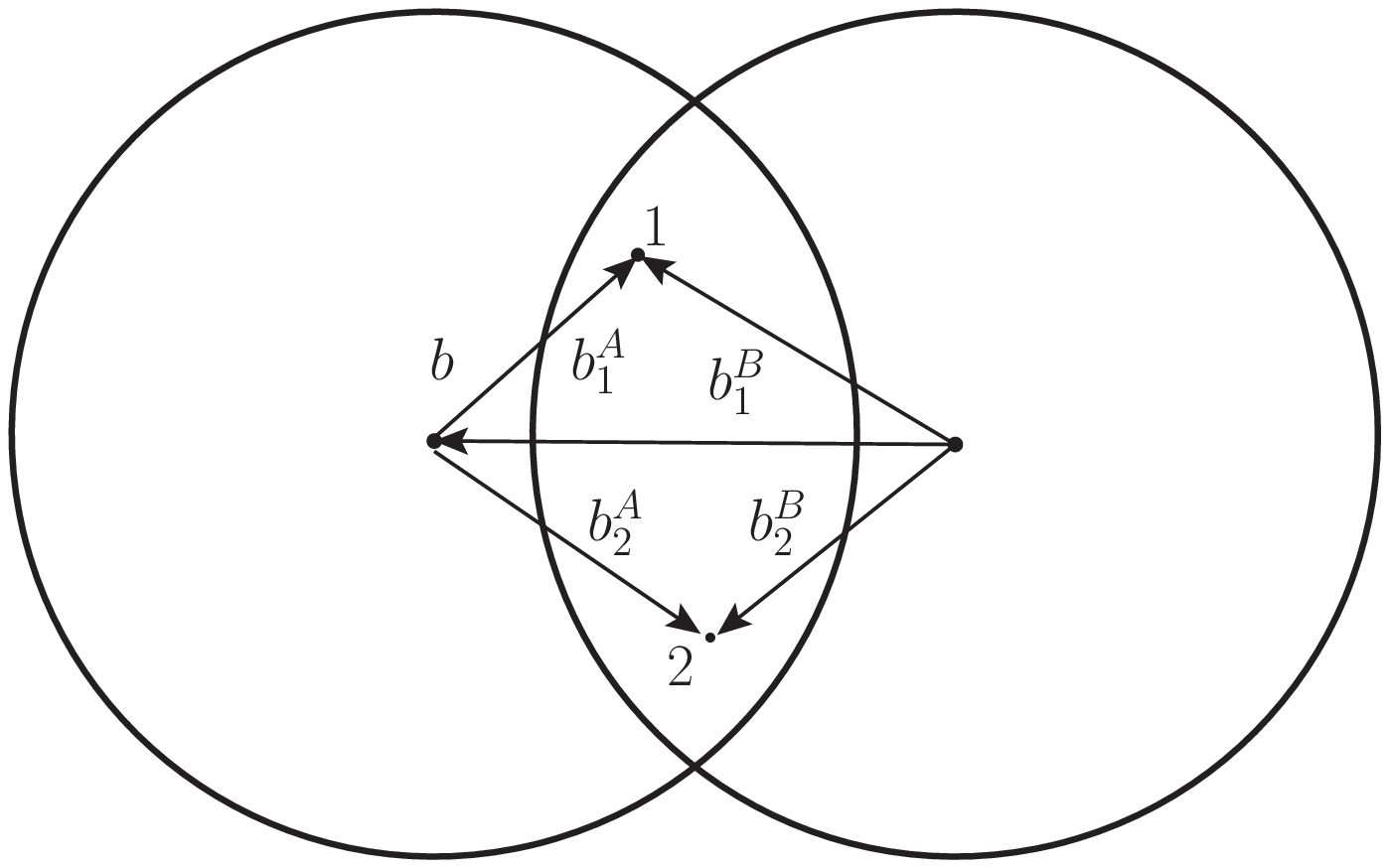}\hspace{5em}
b) \includegraphics[width=0.24\columnwidth]{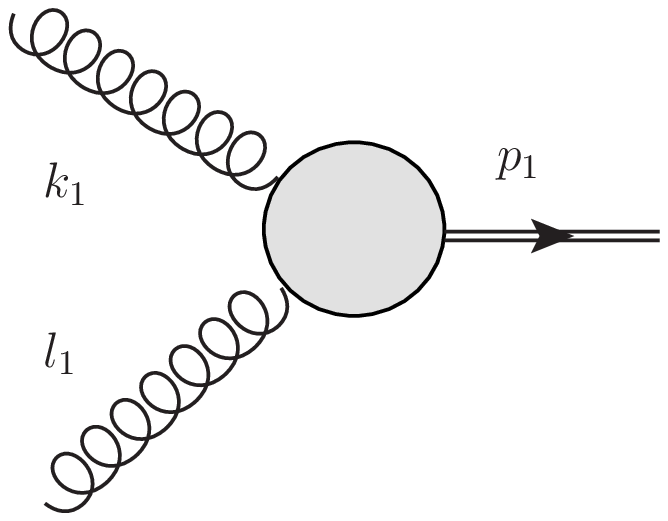}%\hspace{5em}
}
\caption{Colliding hadrons in the impact parameter space (left). The partonic amplitude for the production process $gg\rightarrow V_1$ (right).
\label{fig8}
}
\end{figure}

The relation between $2\mathbb{P}$-p impact factor and unintegrated gluon distributions is read out from the collinear limit of semi-inclusive two particle production process $pp\rightarrow V_1V_2X$ described by the amplitude given in Fig.\ \ref{fig7}. The total cross-section in the impact parameter space is given by the collinear formula (Fig.\ \ref{fig8}a)
\beq
\sigma &=& \hat{\sigma}_1(\alpha_1,\beta_1)\hat{\sigma}_2(\alpha_2,\beta_2)g(\alpha_1,\mu)g(\alpha_2,\mu)g(\beta_1,\mu)g(\beta_2,\mu) \nonumber\\
&&\times \; \int d^2\nvec{b}d^2\nvec{b}^A_1d^2\nvec{b}^A_2 \tilde{S}(\nvec{b}_1^A)\tilde{S}(\nvec{b}-\nvec{b}_1^A)\tilde{S}(\nvec{b}_2^A)\tilde{S}(\nvec{b}-\nvec{b}_2^A),
\eeq
where we assumed that 2-gluons density distribution factorizes into two single gluon distributions $g(\alpha)$ and $\hat{\sigma}_{1,2}$ are cross-sections for the $gg\rightarrow V_{1,2}$ productions at the partonic level. The function $\tilde{S}(\nvec{b})$, which is approximated by the Gaussian form, $S(\nvec{b}) = \exp(-b^2 / R_p ^2) / \pi R_p^2$, is normalized 
$$
\int d^2\nvec{b} \tilde{S}(\nvec{b}) = 1,
$$
where $R_p$ is the effective proton radius. In the momentum space the total cross-section reads
\be
\label{cross_ppVV1}
\sigma = {\hat{\sigma}_1(\alpha_1,\beta_1)\hat{\sigma}_2(\alpha_2,\beta_2) \over \sigma_{\mathrm{eff}}}\,
g(\alpha_1,\mu)g(\alpha_2,\mu)g(\beta_1,\mu)g(\beta_2,\mu),
\ee
where
\be
\sigma^{-1}_{\mathrm{eff}} = 
\int \frac{d^2\nvec{k}}{(2\pi)^2}S^4(\nvec{k} ),\qquad
\tilde{S}(\nvec{b}) = \int\frac{d^2\nvec{k}}{(2\pi)^2}S(\nvec{k} )e^{i\nvec{k} \nvec{b}}, 
\ee
and the normalization condition translates into $S(\nvec{0})=1$.

The partonic amplitude for the production $gg\rightarrow V_1$ has a form (see Fig.\ \ref{fig8}b)
$$
\mathcal{M} = \epsilon_{(\lambda_k)}^{\rho_1}\epsilon_{(\lambda_l)}^{\sigma_1}S^{a_1b_1}_{1\,\rho_1\sigma_1}(k_1,l_1,p_1),
$$
where $S^{a_1b_1}_{1\,\rho_1\sigma_1}$ describes a production amplitude and $\epsilon_{(\lambda)}^{\rho}$ is a gluon polarization vector.
The cross-section, in leading $s$ approximation, reads:
\be
\label{cross_ggV}
\hat{\sigma}_1 = \frac{\pi s}{N_c^2-1}\frac{\alpha_1\beta_1}{\nvec{k}^2_{1\perp}\nvec{l}^2_{1 }}|V_1(k_1,l_1,p_1)|^2\delta(\alpha_1\beta_1s-m^2_{1\perp}),
\ee
where $k_1\approx \alpha_1^kp_A+k_{1\perp}, l_1\approx \beta_1^lp_B+l_{1\perp}, p_1 = \alpha_1p_A+\beta_1p_B+p_{1\perp}$ and
$\alpha_1\approx\alpha_1^k, \beta_1\approx\beta_1^l$. The effective vertex is related to the production amplitude via
$$
V_1(k_1,l_1,p_1)\delta^{a_1b_1} = S^{a_1b_1}_{1\,\rho_1\sigma_1}\frac{p_A^{\rho_1}p_B^{\sigma_1}}{s} .
$$
The amplitude for the process in Fig.\ \ref{fig7} reads
\beq
-i\mathcal{M} &=& \int\frac{d^4k_1}{(2\pi)^4} S_{A\nu_1\nu_2}^{a_1a_2}(p_A,r_A,A^\prime,k_1,k_2)
\frac{d^{\nu_1\rho_1}d^{\nu_2\rho_2}}{k_1^2k_2^2} S_{1\,\rho_1\sigma_1}^{a_1b_1}(k_1,p_1-k_1,p_1) \nonumber\\
&& \times \;  S_{2\,\rho_2\sigma_2}^{a_2b_2}(k_2,p_2-k_2,p_2)\frac{d^{\sigma_1\mu_1}d^{\sigma_2\mu_2}}{(p_1-k_1)^2(p_2-k_2)^2} S_{B\mu_1\mu_2}^{b_1b_2}(p_B,r_B,B^\prime,p_1-k_1,p_2-k_2),
\eeq
and $k_2=p_A-p_{A^\prime}-k_1$. The cross section is then expressed as
\be
\label{cross_ppVV0}
d\sigma_{pp\rightarrow V_1V_2X} = \frac{\bar{|\mathcal{M}|^2}}{4(p_A\cdot p_B)}d\Phi_{pp\rightarrow V_1V_2X},
\;\; \bar{|\mathcal{M}|^2} = \frac{1}{4}\sum_{r_A,r_B}\sum_{A^\prime, B^\prime} |\mathcal{M}|^2 \, ,
\ee
and the phase space integral takes the form
\be
d\Phi_{pp\rightarrow V_1V_2X} = (2\pi)^4\delta (p_A+p_B-p_{A^\prime}-p_{B^\prime}-p_1-p_2) d\Phi_{A^\prime}d\Phi_{B^\prime}d\Phi_{p_1}d\Phi_{p_2} .
\ee
The integrals $d\Phi_{A^\prime}, d\Phi_{B^\prime}$ describe the phase space integrations of $A^\prime$ and $B^\prime$ final states whereas
$d\Phi_{p_1}, d\Phi_{p_2}$ correspond to single-particle phases-spaces for the $V_1, V_2$ particles.

\begin{figure}
\centerline{
\includegraphics[width=0.55\columnwidth]{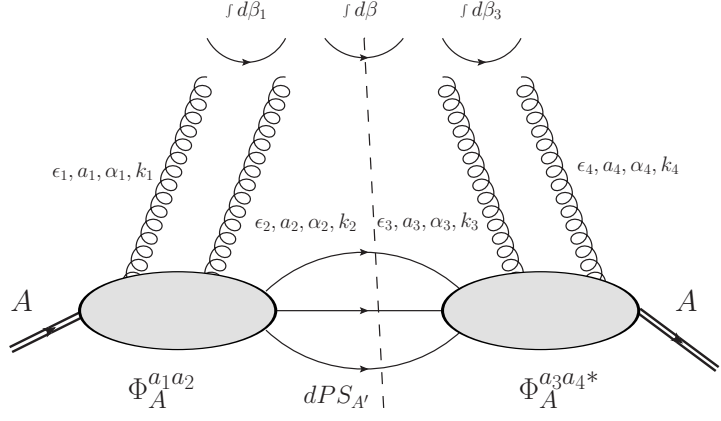}%\hspace{5em}
}
\caption{
The double gluon distribution.
\label{fig9}
}
\end{figure}

Using the appropriate eikonal approximation for gluon polarization tensors and inserting unities in the form
\beq
&&1=\int\frac{d^4k}{(2\pi)^4} (2\pi)^4\delta(k-p_A+p_{A^\prime}),\qquad 1=\int\frac{d^4l}{(2\pi)^4} (2\pi)^4\delta(l-p_B+p_{B^\prime}), \nonumber\\
&&1=\int\frac{d^4l_1}{(2\pi)^4} (2\pi)^4\delta(k_1+l_1-p_1), \qquad 1=\int\frac{d^4l_3}{(2\pi)^4} (2\pi)^4\delta(k_3+l_3-p_1),
\eeq
one can write the cross-section (\ref{cross_ppVV0}) in a factorized form
\beq
\label{cross_ppVV}
d\sigma_{pp\rightarrow V_1V_2X} &=& 
%\nonumber\\
%&\times &
\frac{4s^4}{(2\pi)^{20}}\int
\frac{d^2\nvec{k}_{1}d^2\nvec{k}_{2}d^2\nvec{k}_{3}}{\nvec{k}^2_{1}\nvec{k}^2_{2}\nvec{k}^2_{3}(\nvec{k}_{1 }+\nvec{k}_{2 }-\nvec{k}_{3 })^2}
\frac{d^2\nvec{l}_{1}d^2\nvec{l}_{2}d^2\nvec{l}_{3}}{\nvec{l}^2_{1}\nvec{l}^2_{2}\nvec{l}^2_{3}(\nvec{l}_{1 }+\nvec{l}_{2 }-\nvec{l}_{3 })^2}
\delta(\nvec{k}_{1 }+\nvec{l}_{1 }-\nvec{k}_{3 }-\nvec{l}_{3 })\nonumber\\
&\times &\int d\beta_k\Phi_{4,\,p}^{a_1a_2a_3a_4}(\beta_k,\nvec{k}_{1 },\nvec{k}_{2 },\nvec{k}_{3 },\nvec{k}_{1 }+\nvec{k}_{2 }-\nvec{k}_{3 })
\int d\alpha_l\Phi_{4,\,p}^{a_1a_2a_3a_4}(\alpha_l,\nvec{l}_{1 },\nvec{l}_{2 },\nvec{l}_{3 },\nvec{l}_{1 }+\nvec{l}_{2 }-\nvec{l}_{3 }) \nonumber\\
&\times &\int d\alpha_{1k}d\beta_{1l}S_{1}^{a_1b_1}(\alpha_{1k},\beta_{1l},k_{1\perp},l_{1\perp},p_1)
S_{1}^{a_3b_3\,\ast}(\alpha_{1k},\beta_{1l},k_{3\perp},l_{3\perp},p_1) \nonumber\\
&\times &\int d\alpha_{2k}d\beta_{2l}S_{2}^{a_2b_2}(\alpha_{2k},\beta_{2l},k_{2\perp},l_{2\perp},p_2)
S_{2}^{a_4b_4\,\ast}(\alpha_{2k},\beta_{2l},\nvec{k}_{1 }+\nvec{k}_{2 }-\nvec{k}_{3 },\nvec{l}_{1 }+\nvec{l}_{2 }-\nvec{l}_{3 },p_2) \nonumber\\
&\times &(2\pi)^4\delta(k_1+l_1-p_1)d\Phi_{p_1}\, (2\pi)^4\delta(k_2+l_2-p_2)d\Phi_{p_2} \, ,
\eeq
and $\nvec{k}_{2 }=\nvec{k}_{ }-\nvec{k}_{1 }, \nvec{l}_{2 }=\nvec{l}_{ }-\nvec{l}_{1 }$.
The $2\mathbb{P}$-$p$ impact factors from the third line of equation (\ref{cross_ppVV}) can be decomposed into three color structures each, (see Fig.\ \ref{fig9}),
\beq
\label{decompos}
&&\int d\beta_k\Phi_{4,\,p}^{a_1a_2a_3a_4}(\beta_k,\nvec{k}_{1 },\nvec{k}_{2 },\nvec{k}_{3 },\nvec{k}_{4 })  =
\mathcal{N}\left[\delta^{a_1a_2}\delta^{a_3a_4}f(\nvec{k}_{1 })S(\nvec{k}_{1 }-\nvec{k}_{2 })
f(\nvec{k}_{3 })S(\nvec{k}_{3 }-\nvec{k}_{4 })\right. \nonumber\\
&& \left. + \; \delta^{a_1a_3}\delta^{a_2a_4}f(\nvec{k}_{1 })S(\nvec{k}_{1 }-\nvec{k}_{3 })
f(\nvec{k}_{2 })S(\nvec{k}_{2 }-\nvec{k}_{4 }) + \delta^{a_1a_4}\delta^{a_2a_3}f(\nvec{k}_{1 })S(\nvec{k}_{1 }-\nvec{k}_{4 })
f(\nvec{k}_{2 })S(\nvec{k}_{2 }-\nvec{k}_{3 }) \right]\, , \nonumber 
\eeq
each, and nine color terms are obtained after their multiplication. However, only one of these terms gives the leading contribution, the other ones being suppressed by the color factor ($1/N_c^2$) or the limiting phase space ($p_1\approx p_2$). The normalization constant $\mathcal{N}$ is determined comparing the collinear limit of equation (\ref{cross_ppVV}) to (\ref{cross_ppVV1}).
Indeed, using decomposition (\ref{decompos}) and noting that the last three lines of (\ref{cross_ppVV}) convert into partonic cross-sections (\ref{cross_ggV}), the differential cross-section (\ref{cross_ppVV}) can be written in the form
\beq
&&\frac{d^4\sigma_{pp\rightarrow V_1V_2X}}{d\alpha_1d\beta_1d\alpha_2d\beta_2} = 
\frac{s^2 16(N_c^2-1)^4}{(2\pi)^{20}}\frac{\mathcal{N}^2}{\alpha_1\beta_1\alpha_2\beta_2}
\int\frac{d^2\nvec{k}_{1}}{\nvec{k}^2_{1 }}f(\alpha_1,k_1 ^2)
\int\frac{d^2\nvec{k}_{2}}{k^2_{2 }}f(\alpha_2,k_2 ^2)
\int\frac{d^2\nvec{l}_{1}}{\nvec{l}^2_{1 }}f(\beta_1,l_1 ^2)\nonumber\\
&&\times\int\frac{d^2\nvec{l}_{2}}{\nvec{l}^2_{2 }}
f(\beta_2,l_2 ^2) \hat{\sigma}_1(\alpha_1,\beta_1)\hat{\sigma}_2(\alpha_2,\beta_2)
\int \frac{d^2\nvec{k}}{(2\pi)^2}S^4(\nvec{k} ) .
\eeq
Then it follows that
$$
\mathcal{N}=\frac{(2\pi)^7}{s(N_c^2-1)^2},
$$
where the general correspondence between the unintegrated $f(\alpha,\nvec{k}^2)$ and collinear $g(\alpha,\mu)$ gluon distributions
$$
\int^{\mu^2} \frac{d^2\nvec{k}}{\nvec{k}^2_{ }}f(\alpha,k^2) = \pi \alpha g(\alpha,\mu)$$ was used.

\end{document}